\documentclass[conference]{IEEEtran}

\ifCLASSINFOpdf
 \usepackage[pdftex]{graphicx}

\else

 \usepackage[dvips]{graphicx}

\fi

\usepackage[cmex10]{amsmath}

\usepackage{algorithmic}

\usepackage{array}

\usepackage{mdwmath}
\usepackage{mdwtab}

\usepackage{amsfonts,amssymb,amsmath,amsthm}
\usepackage{latexsym}
\usepackage{epsfig}
\usepackage{cite}

\usepackage[tight,footnotesize]{subfigure}
\usepackage{fixltx2e}

\newtheorem{thm}{Theorem}

\newtheorem{assumption}[thm]{Assumption}

\newtheorem{pr}[thm]{Proposition}

\newtheorem{definition}[thm]{Definition}

\newtheorem{example}{Example}

\newtheorem{remark}{Remark}

{\begin{list}{}{\leftmargin #1cm\setlength{\labelwidth}{\leftmargin}}}%
{\end{list}}

\newcommand{\bi}{\begin{itemize}}
\newcommand{\ei}{\end{itemize}}
\newcommand{\ben}{\begin{enumerate}}
\newcommand{\een}{\end{enumerate}}
\newcommand{\beq}{\begin{equation}}
\newcommand{\eeq}{\end{equation}}

\IEEEoverridecommandlockouts

\usepackage{url}

\begin{document}
\title{Algebraic Constructions of Graph-Based Nested
    Codes from Protographs}

\author{\IEEEauthorblockN{Christine A.~Kelley}
\IEEEauthorblockA{Department of Mathematics\\
University of Nebraska-Lincoln\\
Lincoln, NE 68588, USA\\
Email:  ckelley2@math.unl.edu}
\and
\IEEEauthorblockN{J{\"o}rg Kliewer}
\IEEEauthorblockA{Klipsch School of Electrical and Computer Engineering \\
New Mexico State University\\
Las Cruces, NM 88003, USA\\
Email:  jkliewer@nmsu.edu}
\thanks{This work has been supported in part by NSF grant CCF-0830666
and in part by NSF grant EPS-0701892.}}

\IEEEaftertitletext{\vspace{-1.6\baselineskip}}
\maketitle

\begin{abstract}
  Nested codes have been employed in a large number of communication
  applications as a specific case of superposition codes, for example
  to implement binning schemes in the presence of noise, in joint
  network-channel coding, or in physical-layer secrecy. Whereas nested
  lattice codes have been proposed recently for continuous-input
  channels, in this paper we focus on the construction of nested
  linear codes for joint channel-network coding problems based on
  algebraic protograph LDPC codes. In particular, over the past few
  years several constructions of codes have been proposed that are
  based on random lifts of suitably chosen base graphs.  More
  recently, an algebraic analog of this approach was introduced using
  the theory of voltage graphs.  In this paper we illustrate how these
  methods can be used in the construction of nested codes from
  algebraic lifts of graphs.
\end{abstract}


\section{Introduction}
Nested codes  have been widely used to implement binning schemes
based on coset codes in the presence of noise for numerous scenarios, for
example for the noisy Wyner-Ziv problem \cite{WZ76} and the dual
Gel'fand-Pinsker problem \cite{GP80}. In particular, for the case with
continuous-input channels, binning schemes based on nested lattice
codes have been proposed in \cite{ZSE02}. Recently, in \cite{WM09} the
authors consider discrete-input channels and present compound LDGM/LDPC constructions
which are optimal under ML decoding.

While nested codes in these contexts are related to joint
source-channel coding problems, the class of algebraic nested codes we
will address in this paper are defined based on a \emph{joint channel
  and network coding} scenario. Such nested codes have been originally
proposed in \cite{XFKC07a} for the generalized broadcast relay
problem, where a relay node broadcasts $N$ packets to several
destination nodes, which already know some of the packets \emph{a
  priori}. A related concept was used in \cite{HH06} in the context of
two-way relaying. The idea is that instead of information words,
codewords of different subcodes ${\cal C}_{\ell}$, $1\le \ell \le N$,
are algebraically superimposed via a bitwise XOR. In contrast to
nested codes for the joint source-channel coding scenario described
above, here each subcode and any arbitrary combination of the subcodes
is intended to form a good \emph{channel} code. In particular, this
also holds for the linear combination of all subcodes, the
global code ${\cal C}$. It has been shown in \cite{KDH07} for a
broadcast scenario with side information that such a construction  is able to
outperform a scheme based on a separation of channel and network
coding for non-ergodic discrete-input fading channels. In these
applications we require the subcodes to be better in threshold and/or
in error-floor than the global code.

In this paper we focus on array-code type constructions
\cite{Fan00,TSSFC04} and propose an algebraic design of nested linear
codes based on protograph LDPC codes \cite{Tho03,DDTJ05}. In
particular, in \cite{Kel08,KW08} a lifting technique based on voltage
graphs has been proposed which has been shown to provide a large girth
of the code graph and thus a good error-floor performance. In contrast
to previous approaches based on concatenated and random LDGM codes
\cite{XFKC07a,XFKC07} and also to constructions based on random LDPC
codes we show that the advantage of the above algebraic constructions
in the error floor regime also carries over to the nested code
setting.


\section{Preliminaries}

\subsection{Nested codes}
Consider $M$ different information vectors $i_\ell$ of length
$K_\ell$, $\ell=1,\dots,M$, which we want to encode jointly in such a
way that each information vector is associated with a codeword from a
different subcode.  The overall codeword $c$ is generated by
multiplying the concatenation of all information vectors with a
generator matrix $G$ of the global code ${\cal C}$ according to
\begin{multline}
c^T =[i_1^T, \, i_2^T\, \dots
i_M^T]
\begin{bmatrix}
G_1\\
\vdots\\
G_M
\end{bmatrix}=[i_1^T, \, i_2^T\, \dots
i_M^T] G = \\{i}_1^T\,G_1\oplus
{i}_2^T\,G_2 \oplus \dots \oplus
{i}_M^T\,G_M,
\label{eq:nested1}
\end{multline}
where each of the subcodes ${\cal C}_\ell$ with generator $G_\ell$ of
rate $R_\ell=K_\ell/N$ is associated with the corresponding
information vector $i_\ell$ and $\oplus$ represents a bitwise XOR. The
goal is now to find general systematic design strategies where the
subcodes, any combination of subcodes, and the global code ${\cal C}$
have good threshold and/or error floor properties.

For the sake of simplicity we focus on $M=2$ and the binary case in the
following. Our aim is to design an LDPC code such that its generator matrix $G$
satisfies \eqref{eq:nested1}, where $H\in \{0, 1\}^{(N-K_1 - K_2)\times N}$
represents a corresponding  parity check matrix. If $G$ is
not rank deficient, the null space of $H$ of dimension $(N-K_1-K_2)$
contains the codewords $c_1^T \oplus c_2^T = i_1^T G_1 \oplus i_2^TG_2$.

Likewise, the columns of the parity check matrices $H_1$, $H_2$
associated with $G_1$, $G_2$ each form a basis for their null spaces
of dimensions $(N-K_1)$ and $(N-K_2)$, respectively. A necessary
condition to prevent $G$ from having a rank smaller than $K_1+K_2$ is that
$H_1$, $H_2$ cannot have more than $(N-K_1-K_2)$ linear independent
parity check equations in common.  Based on these considerations, we
propose the following design strategy. First, randomly generate a
matrix $M\in \{0, 1\}^{N\times N}$ of full rank $N$, according to a
given row and column degree distribution. This matrix is then
partitioned into three submatrices
\begin{equation}
M^{(N\times N)} =\Big{[}M_1^{(N\times K_2)} \ M_2^{(N\times K_1)} \
M_3^{(N\times (N-K_1-K_2))} \Big{]}^T.
\label{eq:M}
\end{equation}

Next, the individual parity check matrices for the nested code are
obtained as
\begin{gather*}
H = [M_3^{(N\times (N-K_1-K_2))}]^T, \\
H_1^{((N-K_1) \times N )} = \Big{[} M_1^{N\times K_2}  \ M_3^{(N\times (N-K_1-K_2))} \Big{]}^T,\\
H_2^{((N-K_2) \times N )}  = \Big{[} M_2^{N\times K_1}  \
M_3^{(N\times (N-K_1-K_2))} \Big{]} ^T.
\end{gather*}

Thus, both $H_1$ and $H_2$ are guaranteed to have a null space of
dimensions $(N-K_1)$ and $(N-K_2)$, respectively, and $H$ has $(N
-K_1-K_2)$ parity check equations that are satisfied by ${\cal C}_1$ and
${\cal C}_2$.
\vspace{-0.5ex}
\begin{pr}
The nested code property in \eqref{eq:nested1} holds also if
$M$ and thus one or more of the matrices $H$, $H_1$, and $H_2$ are
(row) rank deficient. For a rank deficit $r$ of the   check matrix $H$
the rate loss for the global code ${\cal C}$ is given as $\Delta R\le r/N$.
\end{pr}
\vspace{-0.5ex}
\begin{proof}
 Denote the rank deficit for the matrices $M_1$, $M_2$ as
  $r_1\ge 0$, $r_2\ge 0$, respectively.
This means that $G_1$ has now a rank of at least
$K_1+r+r_1$, and $G_2$ a rank of at least $K_2+r+r_2$, resp.,  which leads to an
overall rank of at least $K_1+K_2+r+r_1+r_2$ for the generator
matrix $G$. Since both subcodes have  at most $N-K_1-K_2-r$ check equations in
common the row rank of $G$ must not be smaller than $K_1+K_2+r$ to
ensure the nested code property which is satisfied for any $r_1\ge 0$,
$r_2\ge 0$. By setting $R_1'+R_2'=(K_1+K_2+r)/N$
where $R'_1$ and $R'_2$ denote the new rates for the subcodes ${\cal
  C}_1$ and ${\cal C}_2$, a
rate loss of  $\Delta R\le r/N$ for the code ${\cal C}$ is obtained.
\end{proof}
Note that an extension of the above design strategy to $M>2$ can be obtained
in a straightforward way by modifying the partitioning and construction of
$M$ in \eqref{eq:M}.

\subsection{Voltage graphs}

%

\begin{figure}
\centering{\resizebox{2.5in}{1.3in}{\includegraphics{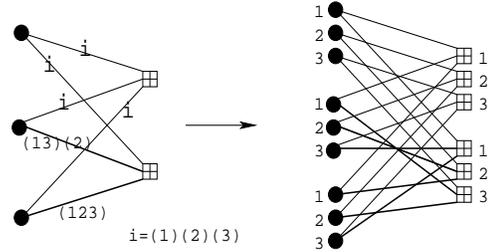}}}
\vspace{-1ex}
\caption{A permutation voltage graph $\mathcal{G}$ is shown on the
  left and its derived graph $\hat{\mathcal{G}}$ on the right, where
  $\mathcal{G}$ serves as a protograph for $\hat{\mathcal{G}}$. The
  darker edges correspond to the connections between the clouds of
  vertices incident with the nontrivial labeled edges.}
\vspace{-4ex}
\label{figure1}
\end{figure}

An algebraic construction of specific covering spaces for graphs was
introduced by Gross and Tucker in the 1970s \cite{GT87}.  For a graph
$\mathcal{G}=(V_{\mathcal{G}},E_{\mathcal{G}})$, a function $\alpha$
called an {\em ordinary voltage assignment}, maps the positively
oriented edges to elements from a chosen finite group ${\sf G}$,
called the voltage group. Each edge in $\mathcal{G}$ has a positive
and negative orientation, and the negative orientation is assigned the
inverse group element.  The base graph $\mathcal{G}$ is called an {\em
  ordinary voltage graph}. The values of $\alpha$ on the edges are
called {\em voltages}.  A new graph $\mathcal{G}^{\alpha}$, called the
{\em (right) derived graph}, is a $|{\sf G}|$-degree lift of
$\mathcal{G}$ and has vertex set $V \times {\sf G}$ and edge set $E
\times {\sf G}$, where if $(u,v)$ is a positively oriented edge in
$\mathcal{G}$ with voltage $b$, then $(u,a)$ is connected to $(v, ab)$
in $\mathcal{G}^{\alpha}$.  Alternatively, another construction takes
the voltage group to be the symmetric group $S_n$ on $n$ elements and
has $\alpha$ map the positively-oriented edges of $\mathcal{G}$ into
$S_n$. This yields a {\em permutation voltage graph}. The {\em
  permutation derived graph} $\mathcal{G}^{\alpha}$ is a degree $n$
lift (instead of $n!$) with vertices $V \times \{1,\ldots,n\}$ and
edges $E \times \{1,\ldots,n\}$.  If $\pi \in S_n$ is a
permutation voltage on the edge $e = (u, v)$ of ${\mathcal{G}}$, then
there is an edge from $(u, i)$ to $(v, \pi(i))$ in
${\mathcal{G}}^{\alpha}$ for $i = 1,2,\ldots,n$.
We will represent
each vertex $(v, i)$ and each edge $(e,i)$ in the derived graph by
$v_i$ and $e_i$, respectively. 
In both cases, the labeled base graph (i.e. voltage graph) algebraically determines a specific lift of the graph.
Fig.~\ref{figure1} shows a permutation voltage graph $\mathcal{G} =
K_{2,3}$ with two nontrivial permutation voltages on its edges to the
group $S_3$, and the corresponding degree 3 permutation derived graph.

Henceforth, derived (lifted) graphs will be denoted by
$\hat{\mathcal{G}}$ since the voltage assignment $\alpha$ should be
clear from context. In this paper we will focus on permutation voltage
graphs for designing nested codes.


\section{Nested codes from protographs}
\label{sec:nested}

We now describe a simple method to construct nested codes from
protographs in which the base Tanner graphs corresponding to
small parity-check matrices $H_1, H_2$, and $H$ are lifted to obtain Tanner
graphs with corresponding parity-check matrices $\hat{H}_1, \hat{H}_2$,
and $\hat{H}$. The simplicity of our method is that it involves just one
lifting of the base graph $\mathcal{G}_M$ corresponding to $M$.

We start with a small bipartite  base graph $\mathcal{G}_M$  with $n$ left vertices, denoted by the set $L$, and $n$ right vertices, denoted by the set $R$.  The matrix $M$ is the incidence matrix of the graph $\mathcal{G}_M$. The $n$ left vertices are the variable nodes  and the right nodes are the constraint nodes (parity-check nodes) of the base graph. We partition the set of right nodes $R$ in $\mathcal{G}_M$ into three disjoint subsets $S_1, S_2$ and $T$ of sizes $k_1, k_2$ and $n-k_1-k_2$, respectively, i.e., $S_1 \cup T \cup S_2 = R$. We define the base graphs for the matrices $M_1, M_2$, and $M_3$  as follows:
\begin{itemize}
\item Let $\mathcal{G}$ denote the induced subgraph of $T$ in
  $\mathcal{G}_M$. Note that $\mathcal{G}$ is a bipartite graph with
  $n$ left vertices of $L$  and $(n-k_1-k_2)$ vertices of $T$.  The
  corresponding parity-check matrix of $\mathcal{G}$ is $M_3$. Lifting
  $\mathcal{G}$ by a degree $m$ lift gives  the derived graph
  $\hat{\mathcal{G}}$ with corresponding parity-check matrix
  $\hat{H}$ for the code ${\cal C}$. The size of $\hat{H}$ is $m(n-k_1-k_2) \times mn$.

\item Let $\mathcal{G}_1$ denote the induced subgraph of $ S_2\cup T$
  in $\mathcal{G}_M$. Note that $\mathcal{G}_1$ is a bipartite graph
  with $n$ left vertices of $L$  and $(n-k_1)$ vertices of $S_2\cup
  T$.  The corresponding parity-check matrix of $\mathcal{G}_1$ is the
  matrix $H_1$. Lifting   $\mathcal{G}_1$  by a degree $m$ lift  gives
  the derived graph $\hat{\mathcal{G}}_1$ with corresponding
  parity-check matrix  $\hat{H}_1$ for the first subcode ${\cal C}_1$. The size of $\hat{H}_1$ is $m(n-k_1) \times mn$.

\item Similarly, let $\mathcal{G}_2$ denote the induced subgraph of $
  S_1\cup T$ in $\mathcal{G}_M$. Note that $\mathcal{G}_2$ is a
  bipartite graph with $n$ left vertices of $L$  and $(n-k_2)$
  vertices of $S_1\cup T$.  The corresponding parity-check matrix of
  $\mathcal{G}_2$ is the matrix $H_2$. Lifting   $\mathcal{G}_2$  by a
  degree $m$ lift  gives  the derived graph $\hat{\mathcal{G}}_2$ with
  corresponding parity-check matrix  $\hat{H}_2$ for the second
  subcode ${\cal C}_2$. The size of $\hat{H}_2$ is $m(n-k_2) \times mn$.

\item The lifts of each of the three graphs $\mathcal{G}_1$, $\mathcal{G}_2$, $\mathcal{G}$ can be done simultaneously by simply lifting the base graph $\mathcal{G}_M$ by a degree $m$ lift in an appropriate way.

\end{itemize}

The blocklength of the lifted nested code is $N =nm$ and the
dimensions of the lifted subcodes are $K_1 \ge k_1m$ and $K_2 \ge
k_2m$ with equality if and only if $\hat{H}_1$ and $\hat{H}_2$ are not
rank deficient.
This construction approach can be extended to nested codes having more
than two component codes in a straightforward way.


With the method outlined above, the design problem of the nested codes
reduces to finding a suitable assignment of permutations (or, more
generally, group elements) to the edges of the base graph
$\mathcal{G}_M$.  Using random permutations is one avenue, however, we are interested in
permutations that are
determined algebraically to obtain an algebraic construction.

In the following we focus on irregular constructions since
 by starting from a regular $(d_v,d_c)$ code ${\cal C}$ with
variable node degree $d_v$ and check node degree $d_c$ the
corresponding subcodes will be regular $(d_v+c,d_c)$ codes with $c>0$.
For the binary-input AWGN channel this typically leads to subcodes
with larger thresholds \cite{AKB04} than the code ${\cal C}$, which
is not desired. By using irregular constructions for the nested code
we can keep a certain fraction of degree-two variable nodes in
the code to improve the threshold, in particular for the subcodes.

\section{Lifted nested codes using commuting permutations}
In our first construction we combine a variant of the algebraic
construction of LDPC codes presented in \cite{TSSFC04} with the
lifting technique described in Section~\ref{sec:nested} to obtain a family of
quasi-cyclic nested codes. For an integer $m$, the subset of integers of the set $\{0, 1, 2,
\dots,m-1\}$ that are co-prime to $m$ forms a multiplicative group $\mathbb{Z}_m^*$.   (If $m$ is prime, then the set $\{0,1, \dots, m-1\}$ form a Galois field and all the non-zero elements in this set form a multiplicative group.) Let $a$ and $b$ be two
non-zero elements in this multiplicative group with multiplicative orders $o(a)=k$ and
$o(b)=j$, respectively.  For $j<k$, we form the following $j\times k$
matrix $P$ with elements from $\mathbb{Z}_m^*$ that has as its $(s,t)th$
element $P_{s,t}=b^sa^t$ as follows:
\[ {\scriptsize P=\left[ \begin{array}{ccccc}
	1 & a& a^2 & \dots &a^{k-1}\\
	b & ab&a^2b&\dots &a^{k-1}b\\
	\dots&\dots&\dots&\dots&\dots\\
	b^{j-1}&ab^{j-1}&a^2b^{j-1}&\dots&a^{k-1}b^{j-1}\end{array}\right].}
\]

Let $P'$ be any $j\times j$ submatrix of $P$.
Let $\mathcal{G}_M$ be the complete bipartite graph $K_{j,j}$ on $j$ variable nodes $\{v_0,v_1,\ldots,v_{j-1}\}$ and $j$ check nodes $\{r_0,r_1,\dots,r_{j-1}\}$. Let $f(\cdot)$ denote a function mapping the elements in $\{0,1,\dots,m-1\}$ to the set of permutations in the symmetric group $S_{m}$, i.e., set of permutations on  $m$ elements. Specifically, we let $f(x)$ denote the permutation that maps $i\mapsto x+i \mbox{ mod } m$, for $i=0,1,\dots, m-1$.
We assign the permutation $f(P'_{s,t})$ for the edge $(r_s,v_t)$ in
$\mathcal{G}_M$ and lift the graph along with their permutation
labeled edges by a degree $m$ lift. We choose three disjoint subsets
$S_1, S_2,$ and $T$ of the set of check nodes $\{ r_0,r_1, \dots,
r_{j-1}\}$ and obtain the induced graphs $\mathcal{G}_1$,
$\mathcal{G}_2$, and $\mathcal{G}$, as described in Section~3. The
resulting derived (lifted) graph $\hat{\mathcal{G}}_M$ also yields the
lifted graphs $\hat{\mathcal{G}}_1$, $\hat{\mathcal{G}}_2$ and
$\hat{\mathcal{G}}$ and the corresponding parity-check matrices
$H_1,H_2$, and $H$ of the nested code. In particular, the matrix $M$
is the all-ones matrix of size $j\times j$. The corresponding
incidence matrix $\hat{M}$ for the lifted graph $\mathcal{G}_M$ is a
matrix that is a $j\times j$ array of shifted identity matrices, with
the shifts corresponding to the entries in the matrix $P'$. For
example, if the first $j$ columns and $j$ rows of $P$ form the matrix
$P'$, then
\[
{\scriptsize \hat{M}^T=\left[ \begin{array}{ccccc}
 	I_1 & I_a& I_{a^2} & \dots &I_{a^{j-1}}\\
 	I_b & I_{ab}&I_{a^2b}&\dots &I_{a^{j-1}b}\\
 	\dots&\dots&\dots&\dots&\dots\\
 	I_{b^{j-1}}&I_{ab^{j-1}}&I_{a^2b^{j-1}}&\dots&I_{a^{j-1}b^{j-1}}
       \end{array} \right],
} \]
where $I_x$ denotes the $m\times m$ identity matrix cyclically shifted
to the left by $x$ positions.  In a more general array construction in \cite{Fan00}, the shifts in the above construction are chosen randomly from the set $\{ 0, 1, \dots, m-1 \}$.

The base graph $\mathcal{G}_M$ may be viewed as a permutation voltage
graph, and its $m$-degree lift $\hat{\mathcal{G}}_M$ as a permutation
derived graph, where the local voltage group consists
of the permutations  that map $i\mapsto x+i \mbox{ mod } m$, for $i=0,\ldots,m-1$, where $x$ can take values in $\{0,1, \ldots,m-1\}$.\\

%

The constructed codes are quasi-cyclic and thus have an encoding
complexity of $O(1)$ per symbol \cite{LCZLF06}.
The codes have performance comparable to random LDPC
codes for short to moderate blocklengths. However, at large block
lengths, the random codes are expected to outperform this construction
as the distance and girth of these codes are limited. Specifically,
whenever there is a $K_{2,3}$ subgraph in the base graph, the girth of
the lifted nested codes is at most 12, and the distance is limited by
$(j+1)!$ for a column weight $j$ parity-check matrix \cite{TSSFC04,
  KW08, MD99}. These limitations motivate the use of non-commuting
voltages in the
construction in the next section to help
surpass these girth and distance limitations.



\section{Lifted nested codes using noncommuting permutations}
In our second construction we combine the algebraic construction of
LDPC codes presented in \cite{Kel08} with the lifting technique
described in Section~3 to obtain a family of nested codes from lifts
using nonabelian voltage groups.  When the permutations
assigned are pairwise non-commuting and meet the cycle structure and
connectivity requirements as outlined in \cite{Kel08},
the derived graphs for the nested code and its subcodes
are connected and have improved girth and distance even when the base graph contains a $K_{2,3}$ subgraph.



For an edge $e$, let $e^{-}$ and $e^{+}$ denote the negative and
positive orientations, respectively, of $e$. A {\em walk} in the
ordinary or permutation voltage graph $\mathcal{G}$ may be represented
by the sequence of oriented edges in the order they are traversed,
e.g.  $W=e_1^{\sigma_1}e_2^{\sigma_2}\dots e_n^{\sigma_n}$ where each
$\sigma_i$ is $+$ or $-$ and $e_1,\dots,e_n$ are edges in
$\mathcal{G}$. In this setting, the {\em net voltage} of the walk $W$
is defined as the voltage group product $
\alpha(e_1^{\sigma_1})\alpha(e_2^{\sigma_2})\dots
\alpha(e_n^{\sigma_n})$ of the voltages on the edges of $W$ in the
order and direction of the walk. We now have the following theorem \cite{GT87}.
\begin{thm}
  Let $C$ be a $k$-cycle in the base graph of a
  permutation voltage graph with net voltage $\pi$, and let
  $(c_1,c_2,\ldots, c_n)$ be the cycle structure of $\pi$. Then the
  pre-image of $C$ in the derived graph has $c_1+c_2+\cdots+c_n$
  components, including, for each $j = 1,\ldots,n$, exactly $c_j$
  $kj$-cycles. $\hfill \Box$
\label{grosstucker_thm2}
\end{thm}
Here we distinguish between a $k$-cycle in a graph which is a closed walk 
of length $k$, and a cycle of a permutation which is a closed set of 
numbers in the cycle representation of the permutation. The {\em cycle 
structure} of a permutation in $S_n$ is a vector 
$(c_1,\ldots, c_n)$ where $c_j$ denotes the number of $j$-cycles in the 
cycle decomposition of the permutation.


We choose permutation voltages that do not have fixed points, and in
fact, do not contain cycles of length $\le 3$. This allows our
construction to surpass the girth 12 restriction that exists in the
abelian case, provided that there are no short products of these
voltages that yield permutations with cycles of size $\le 3$ in their
decomposition. We also choose a voltage group where the only
group element with fixed points is the identity permutation. This
eliminates fixed points in the net voltages of all graph cycles that do
not have the identity permutation as a net voltage. Moreover, ${\sf G}$
has just one orbit when acting on $\{1,2,\ldots,m\}$ so we will assign
permutations to the base graph that generate ${\sf G}$ to meet the
connectivity condition \cite{Kel08}.




We adapt the approach from \cite{Kel08} to determine the permutation
voltage assignment to the edges of $\mathcal{G}_M$.  We choose $m =
pq$ such that $p$ and $q$ are prime, $q < p$, and $q | (p-1)$.  We
construct the nonabelian group ${\sf G}$ of order $m=pq$ generated by
elements $c$ and $d$ such that the order of $c$ is $p$, the order of
$d$ is $q$, and $dc = c^sd$, where $s \not \equiv 1 (\mbox{mod } p)$
and $s^q \equiv 1 (\mbox{mod } p)$.
Further, we construct the
permutation group isomorphic to ${\sf G}$ to use as our permutation
voltage group, which we will also denote by ${\sf G}$.

We form the following $j \times k$ matrix $P$ with $j \le k$ and entries 
in ${\sf G}$ as follow. All the entries on $0$th row and the $0$th column 
of $P$ are assigned the identity permutation. The 0th row and 0th column 
of $P$ correspond to a spanning tree in the base graph $K_{j,k}$. The 
group ${\sf G}$ has one subgroup of order $p$ of the form $\{ 1, c, 
c^2 , \dots, c^{p-1}\}$ and $p$ subgroups of order $q$ of the form $\{1, 
c^id, (c^id)^2, \dots, (c^id)^{q-1}\}$, for $i=0,1,\dots,p-1$. For the 
remaining entries in $P$, we assign non-identity permutations, that are 
mostly chosen from distinct subgroups of ${\sf G}$. If $(j-1)(k-1)\le 
p+1$, (or in general, the number of edges outside the spanning tree is 
at most $p+1$), then there are enough distinct subgroups from which to 
choose the 
permutations. Finally, we ensure that the permutations chosen in $P$ 
generate the group ${\sf G}$. 


%

Let $P'$ be any $j\times j$ sub-matrix of $P$. Then, following the approach in Section~4, the
resulting derived (lifted) graph $\hat{\mathcal{G}}_M$ yields the
lifted graphs $\hat{\mathcal{G}}_1$, $\hat{\mathcal{G}}_2$ and
$\hat{\mathcal{G}}$ and the corresponding parity-check matrices
$H_1,H_2$, and $H$ of the nested code. In particular, the matrix $M$
is the all-ones matrix of size $j\times j$. The corresponding
incidence matrix $\hat{M}$ for the lifted graph $\mathcal{G}_M$ is a
matrix that is a $j\times j$ array of permutation matrices, with
the permutations corresponding to the entries in the matrix $P'$.





This construction and the one in Section~4 can be adapted to any base
graph with $j$ check nodes and $k$ variable nodes, not just a complete
base graph, by simply replacing the entries corresponding to no edge
connections with all zero matrices. In this way, other degree
distributions can be accommodated, such as in the design example in
Section~6. Other spanning trees can be chosen for the identity permutations, accordingly.
Furthermore, the above construction can be extended in a  natural way even when the
matrix $M^T$ is an  $j'\times j$ matrix for $j'< j$, thereby yielding a
rank deficient matrix $\hat{M}^T$ as described in Proposition 1. The design
example in the next section uses such a matrix.

\section{Design example}

We start with a base graph with 12 check nodes and 16 variable nodes
having the following check to variable incidence  (or, base
parity-check) matrix $M'$:

{\tiny \[
{M'}^T=\left[\begin{array}{cccccccccccccccc}
1&0&0&0&0&0&0&0&1&0&0&1&0&0&0&1\\
0&0&0&0&0&0&0&1&1&0&0&0&0&0&0&1\\
0&1&0&1&0&0&1&0&0&0&1&0&0&1&0&1\\
0&1&1&1&1&0&1&1&1&0&0&0&0&0&0&0\\
1&0&0&0&0&0&0&0&0&1&0&0&1&0&1&1\\
1&0&0&1&0&1&0&0&0&1&0&0&0&1&0&1\\
0&0&0&1&0&1&0&0&1&0&1&1&0&0&1&0\\
1&0&1&0&1&0&0&1&0&0&0&1&1&0&0&0\\
1&0&0&0&0&0&0&1&0&0&1&0&1&1&1&0\\
0&0&1&0&1&1&0&0&1&0&0&1&0&0&0&0\\
0&0&1&0&0&0&0&0&0&0&1&0&1&0&0&1\\
1&0&0&0&0&1&0&0&1&0&0&0&0&0&0&1\end{array}\right]
\]
} The first 10 rows correspond to the base parity-check matrix of the
first subcode ${\cal C}_1$ and the last 10 rows correspond to the base
parity-check matrix of the second subcode ${\cal C}_2$ and rows 3-10
correspond to the base
parity-check matrix of the global code ${\cal C}$. Using the construction
approaches given in Sections~4 and~5, two groups, each of size
$m=305$, are chosen.  As a first step, a $12\times 16$ matrix $M^T$
having all one entries is considered. In the first construction
described in Section~4, the entries in $M^T$ are replaced with shifted
identity matrices (each having size $m\times m$) to obtain a lifted
matrix $\hat{M}^T$ of size $3660\times 4880$. In the second
construction, a non-commutative group of order $m=305$ is
considered, and the entries in $M^T$ are replaced by $m\times m$
permutation matrices corresponding to the permutations as shown in
Section~5.

For each case, a lifted matrix $\hat{M'}^T$ corresponding to the matrix
${M'}^T$ above is obtained by multiplying the $(i,j)^{th}$ block of
shifted identity or permutation matrix in $\hat{M}^T$ by the
$(i,j)^{th}$ entry in ${M'}^T$. The first 10 row blocks represent the
parity check matrix of the first subcode ${\cal C}_1$, the last 10 row
blocks represent the parity check matrix of the second subcode ${\cal
  C}_2$, and the row blocks 3-10 represent the parity-check matrix of
the global code ${\cal C}$. ${\cal C}_1$ and ${\cal C}_2$ have block
length $N=4880$ and code rate $R_1=R_2=0.375$ (and thus exhibit a rate
loss) whereas ${\cal C}$ has block length $N=4880$ and code rate
$R=0.5$. The choice of $M'$ above yields the following degree
distributions and (exact) density evolution thresholds (in $E_b/N_0$)
for the nested codes: a) Code ${\cal C}$:
$\bar{\lambda}_2=\frac{5}{16},  \bar{\lambda}_4=\frac{8}{16}, \bar{\lambda}_4=\frac{3}{16},
\bar{\rho}_5=\frac{2}{8}, \bar{\rho}_6=\frac{6}{8}$, and density
evolution threshold $0.914$\,dB, where $\bar{\lambda}_i$ (resp.
$\bar{\rho}_i$) denotes the fraction of variable (resp. check) nodes
of degree $i$, and b) codes ${\cal C}_1, {\cal C}_2$:
$\bar{\lambda}_2=\frac{4}{16},
\bar{\lambda}_3=\frac{5}{16},\bar{\lambda}_4=\frac{4}{16},
\bar{\lambda}_5=\frac{3}{16}, \bar{\rho}_4=\frac{2}{10},
\bar{\rho}_5=\frac{2}{10},\bar{\rho}_6=\frac{6}{10}$, and density
evolution threshold  $0.688$\,dB.

\begin{figure}[htb]
  \centerline{\includegraphics[scale=0.5]{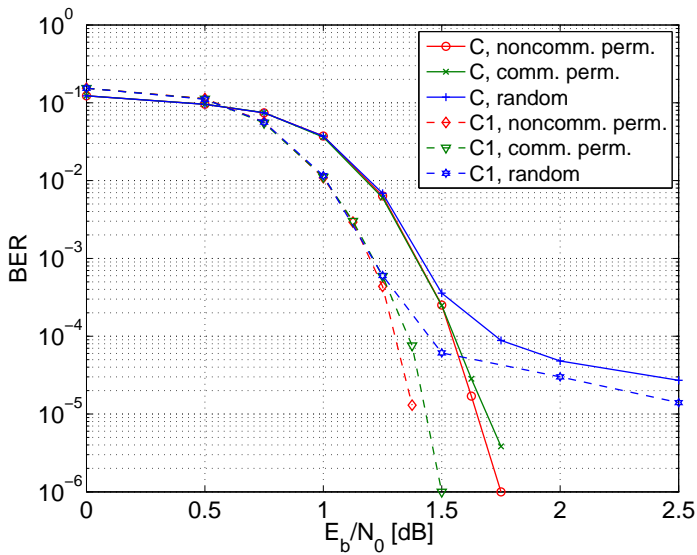}}
  \vspace*{-2ex}
    \caption{BER versus $E_b/N_0$ for the lifted nested codes with
      commuting and noncommuting permutations.}
    \vspace*{-1ex}
    \label{fig:sim}
\end{figure}

Simulation results on the binary-input AWGN channel using belief
propagation decoding are presented in Fig.~\ref{fig:sim} for the
lifted nested code given in above example. (All simulations were run
for a maximum of 50 decoding iterations.  The performance of ${\cal C}_2$ is
almost identical to that of ${\cal C}_1$ and therefore not shown.) The
protograph codes from this paper are compared with randomly designed
protograph codes having identical block lengths, code rates, and degree
distributions in their parity-check matrices.

We can observe from Fig.~\ref{fig:sim} that all subcodes
perform better than the corresponding (overall) codes ${\cal C}$.
Further, the codes obtained from the nonabelian 
group perform  better in the error floor regime than those from the
abelian group, whereas the random constructions are penalized by a significant
error floor due to the low girth of their code graphs.
\section{Conclusions}

In this paper, an algebraic construction of graph-based nested codes is 
introduced. The method relies on a protograph design and a lifting 
technique using algebraic voltage graphs, and may be applied to other 
base graphs with other degree distributions for improved performance. The 
resulting codes have compact description, structure that is well-suited 
for practical implementation in several applications, and a performance 
that is better than that of randomly designed codes.




\bibliographystyle{IEEEtran}

\bibliography{lit}

\end{document}